# PAIRWISE XOR AND XNOR GATES IN SQUEEZED INSTANTANEOUS NOISE-BASED LOGIC

NASIR KENARANGUI [1,§], ARTHUR POWALKA[1], LASZLO B. KISH[1]

[1]*Department of Electrical and Computer Engineering, Texas A&M University, College Station, TX 77843-3128, USA*

Instantaneous noise-based logic (INBL) is a novel computing approach that encodes binary information using stochastic processes. It uses $2M$ orthogonal stochastic reference noises for $M$ noise-bits to construct an exponentially large Hilbert space (hyperspace) of dimension $2^M$. INBL offers a classical alternative to quantum-style parallelism for specific problems with exponential speedup compared to classical algorithms. Building on recent work that introduced pairwise XOR and XNOR operations defined for a symmetric INBL scheme, this paper implements these gates for a squeezed INBL scheme. Hyperspace vectors are product strings corresponding to $M$-bit long binary numbers. The proposed operations can apply pairwise on hyperspace vectors and their superpositions (sums), while remaining compatible with the squeezed reference system. We validate that the squeezed-scheme XOR/XNOR gate operations have correct Boolean behavior over both bitwise and targeted $M$-bit strings and demonstrate that the operations preserve instantaneous evaluation. The results show that the XOR/XNOR toolkit, previously developed for symmetric INBL, can be tailored for the squeezed scheme. This development is a key part of the gate set needed for more complex INBL algorithms in the squeezed INBL scheme and advances the objective of gate universality in INBL. It further strengthens the case for INBL as a flexible, classical computing framework that can emulate some structural advantages of quantum computation.

*Keywords:* instantaneous noise-based logic; squeezed INBL; random telegraph waves; hyperspace vectors; pairwise XOR; pairwise XNOR; superposition computing; classical alternative to quantum computing; parallel gate operations; noise-bits

## 1. Introduction

There has been an ongoing investigation into the advantages demonstrated by current quantum computers, especially when compared with classical Turing machines [1-4]. In this context, "quantum supremacy" is usually taken to mean the existence of carefully chosen computational tasks for which quantum computers are believed to have an exponential speedup compared with solutions on any classical Turing machine. Early proposals such as the original Deutsch–Jozsa algorithm are often cited as paradigmatic examples in this discussion. Instantaneous noise-based logic (INBL) was introduced with the aim of achieving quantum-like exponential speedup for specific problems using classical physical hardware.

§ Corresponding Author.

From a computational perspective, INBL occupies an intermediate position between conventional classical models and fully quantum computation. Classical Turing machines operate on deterministic bit strings with local, sequential updates, whereas quantum Turing machines augment this model with coherent superposition and entanglement over a Hilbert space of dimension $2^M$, subject to unitary evolution and measurement postulates. These distinctions are discussed in more detail in [5].

In contrast, INBL remains a purely classical, stochastic framework, but uses families of orthogonal random telegraph waves to realize a $2^M$-dimensional hyperspace in a single wire, enabling parallel operations on superpositions of hyperspace vectors without invoking quantum coherence. Previous work has shown that this structure can yield quantum-style exponential speedup for specific tasks [6], such as variants of the Deutsch–Jozsa problem and related search or verification problems. Specifically, an INBL search algorithm has been shown to yield exponential speedup even relative to Grover's quantum search for the corresponding task. This further supports the potential of INBL as a classical platform with quantum-style advantages for specific problems [7]. Entire sets of candidate solutions can be encoded and processed simultaneously on a fixed-size physical substrate, rather than being explored one by one on a classical Turing machine. In this sense, INBL leverages classical randomness and orthogonality to emulate some algorithmic advantages usually associated with quantum computation, while remaining within a classical hardware paradigm.

In noise-based logic (NBL) information is carried by stochastic processes. Orthogonal noise-bits serve as base states for computation [8-10]. The reference noise system consists of $2M$ noise sources for $M$ noise bits, and logic operations are implemented using stochastic signals, see Figure 1. $M$ can be viewed as being analogous with the quantum bit resolution in an equivalent quantum computer (the number of qubits). In the correlator approach, states are recognized through cross-correlation with reference noises, which requires time-averaging. INBL is a class of NBL that eliminates time averaging requirement, and operations are completed "instantaneously" without convergence delays [11].

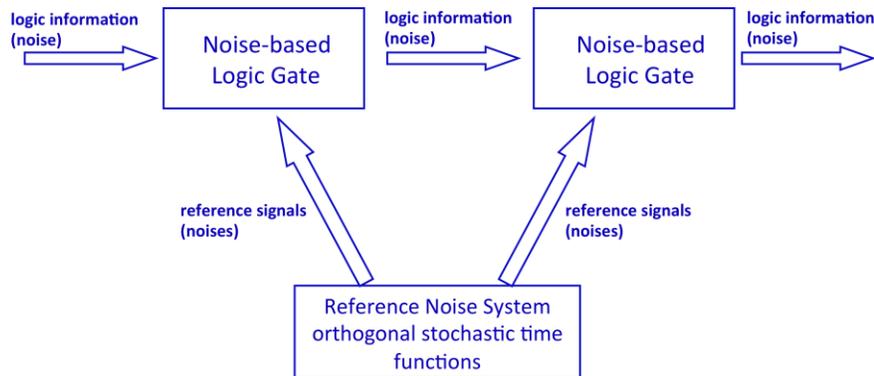

**Figure 1.** General implementation for NBL system. In $M$ noise-bit resolution NBL, $2M$ orthogonal reference noises form the reference noise system [12,13].

For an INBL system with $M$-bit resolution, each noise-bit is represented by two independent random telegraph waves (RTWs), corresponding to the logic low and high states. $2M$ orthogonal reference noises stored in computer memory form the reference system in the INBL logic scheme, see Figure 2 (a). In this sense, the reference RTWs play a role analogous to an orthonormal basis in a quantum $2^M$-dimensional Hilbert space, with orthogonality ensured by their zero cross-correlation and equal-energy statistics. Reference signals associated with the relevant noise-bits are transmitted on the reference wires [8-16]. Finally, the Hilbert space synthesizer generates the string products with reference noises and their superpositions (sums) [8,10,14]. As an example, Figure 2 (b) shows the two-stage structure of the Hilbert space synthesizer [14]. In this example the hyperspace vector $R_{1,0}(t)R_{2,1}(t)$ associated with binary number 01, and the hyperspace vector $R_{1,1}(t)R_{2,0}(t)$ associated with binary number 10 are superimposed $R_{1,0}(t)R_{2,1}(t) + R_{1,1}(t)R_{2,0}(t)$. Here, "superposition" refers to the classical sum of stochastic signals (RTWs) and should not be confused with quantum superposition in a Hilbert space of probability amplitudes. The superposition of these two hyperspace vectors yields a composite noise signal in which the individual noise-bits are no longer separable [7]. Whenever the first bit is high in the product structure, the second is low, and vice versa. This non-factorizable joint structure mimics the logic pattern of the Bell state in quantum computers $(|\,10\rangle + |\,01\rangle)/\sqrt{2}$ [7,17], providing a classical, INBL analogue of entanglement. Similarly, a superposition of the hyperspace vectors corresponding to 00 and 11 exhibits the same correlated bit pattern as the $\Phi$-type Bell states, while a superposition of 01 and 10 captures the anticorrelated pattern of the $\Psi$-type Bell states. In all cases, the analogy is limited to the classical correlation structure of the bit strings; INBL employs real-valued stochastic superpositions and does not reproduce the complex relative phases that distinguish, for example, $\Psi^+$ from $\Psi^-$ or $\Phi^+$ from $\Phi^-$.

The analogy to Bell states used here is purely structural at the level of classical correlation patterns: superpositions of hyperspace vectors can reproduce the same joint bit statistics as the corresponding Bell-state basis patterns (e.g., perfect correlations or anticorrelations between two bits). INBL operates entirely with classical stochastic signals and does not reproduce quantum nonlocality, complex phases, or measurement-induced collapse in the quantum-mechanical sense. What it does provide is a deterministic, gate-level mechanism for selectively amplifying or nulling specific hyperspace components (e.g., by grounding the inverse of a target string), which plays a role analogous to "collapsing" onto a marked outcome in INBL search algorithms. Consequently, these "classical entanglement" analogues cannot be used for intrinsically quantum tasks such as teleportation [18,19] and do not by themselves provide any advantage beyond what is achievable with standard quantum algorithms; they only motivate useful structural parallels between INBL hyperspaces and quantum Hilbert spaces.

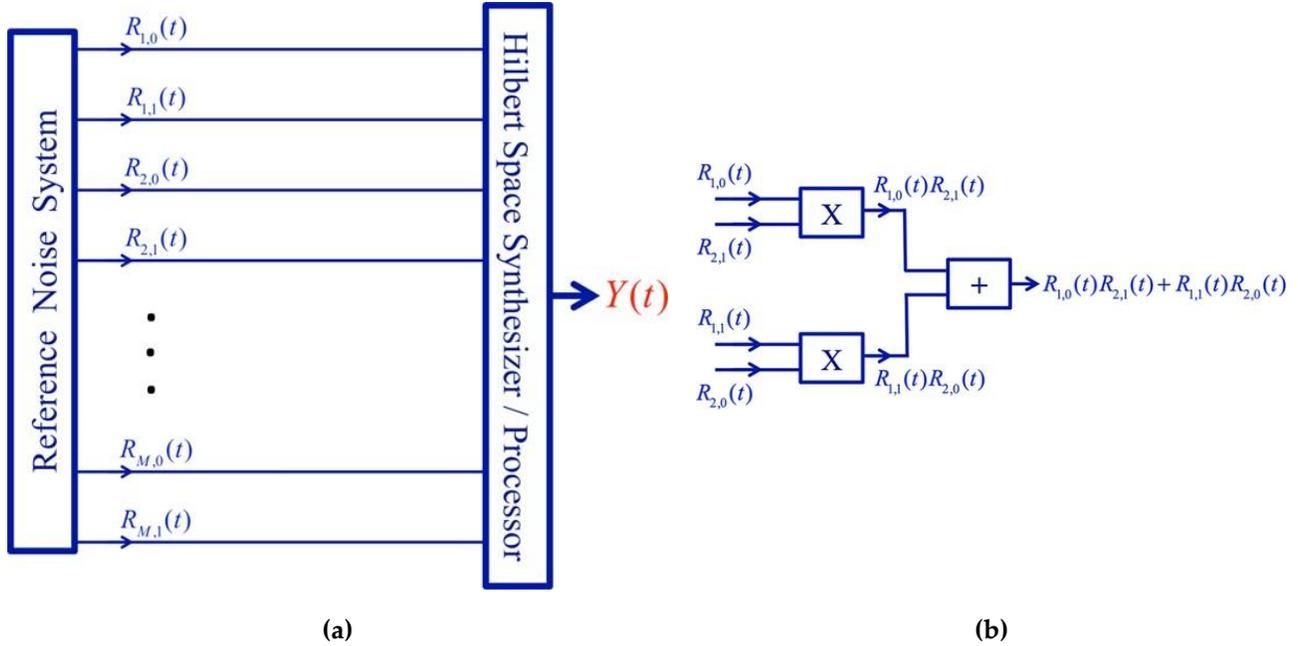

**(a)**                                                    **(b)**

**Figure 2.** Generic INBL processor structure: **(a)** The reference noise system generates and stores the $2M$ references for $M$ noise-bits. Reference wires enable transmission and manipulation of reference noises [13]. For a given superposition state $Y(t)$, Hilbert space synthesizer generates the string products and their superposition.; **(b)** Example of Hilbert space synthesizer showing superposition of hyperspace vectors associated with number 01 and 10 [7].

In general, RTWs can exhibit more than two amplitude levels. RTWs used in the reference noise system, as demonstrated in Figure 3, are discrete-time stochastic signals whose amplitudes flip between $+1$ and $-1$ at each clock cycle with probability $P = 0.5$ [11-14]. In a symmetric INBL scheme, the RTWs representing both low and high states have the same amplitude $\pm 1$.

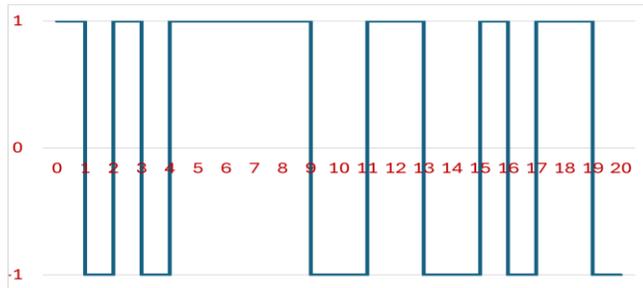

**Figure 3.** The RTWs are square waves with randomly chosen amplitudes of $+1$ or $-1$, with probability of 0.5, during rising edge of each clock cycle.

The essential properties of these RTWs are orthogonality and statistical independence. Two distinct RTWs $R_{i,j}(t)$ and $R_{k,l}(t)$, representing different reference noises, have zero mean and zero cross-correlation [14]:

$$\langle R_{i,j}(t) \rangle = 0, \tag{1}$$

$$\langle R_{i,j}(t) R_{k,l}(t) \rangle = \delta_{i,k}\, \delta_{j,l}. \tag{2}$$

where i ≠ k, j ≠ l, δ is the Kronecker symbol, and ⟨ ⟩ denotes time averaging. In Eq. (2), $\delta_{i,k} = 1$ if $i = k$ and 0 otherwise, and similarly for $\delta_{j,l}$. This captures: Zero mean cross-correlation when $(i, j) \neq (k, l)$, and unit autocorrelation when $(i, j) = (k, l)$. This expression implies that distinct reference RTWs are mutually orthogonal (zero cross-correlation), whereas the same RTW has unit self-correlation, so the set $\{R_{i,j}\}$ forms an orthogonal family of basis waveforms.

RTWs also exhibit a multiplicative property, where the product $R_3(t) = R_1(t)R_2(t)$ yields a new RTW that is uncorrelated with and orthogonal to both $R_1(t)$ and $R_2(t)$. Here, the single-subscript notation $R_1, R_2, R_3$ is used to illustrate three distinct RTWs from the general family $\{R_{i,j}(t)\}$. Additionally, the self-product of an RTW produces a constant "vacuum" vector of ones, $R_1(t)R_1(t) = 1$ [20]. An $M$ bit long product string, composed of the relevant reference noises for their respective noise-bits, is called a hyperspace vector, which corresponds to a decimal number $n$ where $n = 0, 1, \ldots, 2^{M-1}$. Product strings shorter than $M$ noise-bits are also possible, for special purpose applications, but a full hyperspace vector is $M$ bits long.

These properties were leveraged in [21] for pairwise XOR and XNOR operations that were introduced and analyzed for the symmetric INBL scheme. In this approach, the gates are realized through products of RTWs, which support bit-targeted and pairwise operations. This structure allows both hyperspace vectors and their superpositions to be used directly as one of the input pairs.

In the squeezed INBL scheme, the logic low states are represented by constant vectors, while the logic high states are represented by RTWs [15]. "Squeezing" of the low state into a constant simplifies the hardware implementation and subsequent signal processing, because one of the two logic values no longer requires a stochastic source or storage. Either the logic low or logic high reference noises can be mapped to a constant waveform; this choice is conventional and does not affect the underlying gate algorithms. In this work, the logic-low states are assigned the constant value 1, while the logic-high states remain RTWs, following the convention adopted in earlier formulations of squeezed INBL. If instead the high reference noises were squeezed to a constant and the low states realized as RTWs, the same XOR/XNOR constructions would carry through after a relabeling of the reference states, and all results presented here would remain valid up to this notational change. In this paper we present the pairwise XOR and XNOR operations within a squeezed scheme.

## 2. Squeezed INBL System Implementation

In the squeezed INBL scheme, the logic low state is represented by a constant waveform $L(t) = 1$ rather than by a zero signal, see Figure 4. This choice preserves the simplifying benefits of squeezing, only the high state must be generated as a stochastic random telegraph wave. At the same time, it avoids the degeneracy that would arise if any low factor were exactly zero. If the logic low values were mapped to $L(t) = 0$, the presence of a single low bit in a product string would force the entire hyperspace vector to zero, eliminating many distinct product states and severely restricting hyperspace operations. By using $L(t) = 1$, low bits become multiplicatively transparent, and do not require their own noise sources, yet they also do not annihilate the product. As a result, all $M$-bit patterns still correspond to nontrivial hyperspace vectors formed from subsets of high-state RTWs. This allows the squeezed scheme to retain the essential structure of INBL's $2^M$-dimensional product space while reducing hardware complexity and simplifying signal representation [15].

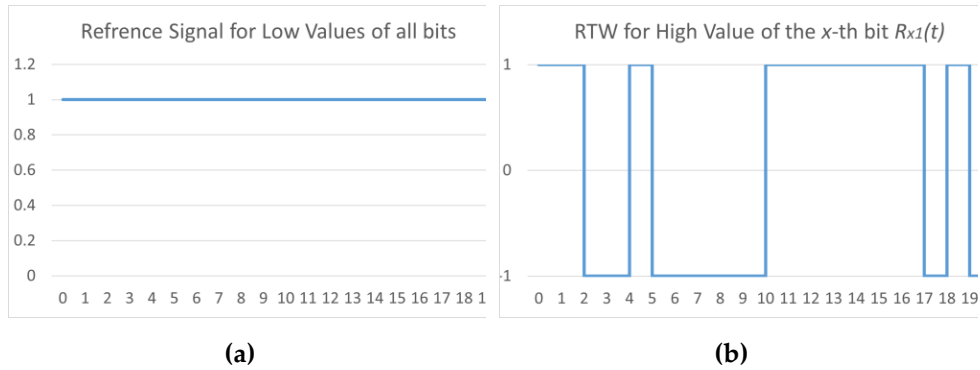

**(a)**    **(b)**

**Figure 4.** Example for the first 20 clock cycles of $x$-th noise-bit with squeezed reference noises. In the squeezed INBL scheme, each noise-bit is still represented by a pair of waveforms: a constant signal corresponding to the logic-low state and an RTW corresponding to the logic-high state. **(a)** For all noise-bits, the low values are represented by a constant reference signal of amplitude $+1$; **(b)** for each noise-bit scheme, high values are assigned a RTW reference noise with amplitude of $+1$ or $-1$ with probability of 0.5.

### 2.1. *Construction of the Reference Noise System*

A pseudorandom number generator (PRNG) can be used to generate RTWs for high reference noises, while low reference noises are equal to constant 1 for each of the $M$ noise-bits [21]. In practice, the RTWs used in the reference noise system can be generated either by algorithmic PRNGs or by hardware sources such as chaotic oscillators and physical white-noise based true random number generators, as widely explored in the cryptography literature [22-24]. For the numerical demonstrations in this work, the RTWs were produced by a standard software PRNG. Different seeds naturally produce different realizations of the RTWs, but the statistical properties relevant for INBL (zero mean, orthogonality, and equal variance) are preserved, and we did not observe any qualitative dependence of the gate behavior on the particular choice of seed.

The reference system, stored in the computer memory, only requires $M$ orthogonal RTWs, which are associated with the high reference noises:

$$R_{11}(t), R_{21}(t), \dots, R_{M1}(t). \tag{3}$$

The minimum period length requirement to reach the target error probability of $0.5^T = 10^{-25}$ is $T = 83$ clock cycles. Where $10^{-25}$ is the theoretical error probability of an idealistic gate in modern computers [25]. For the verification of squeezed INBL operations in this paper the number of RTW clock cycles is chosen to be greater than $100$.

## 2.2. *Generation of Hyperspace Vectors*

Each hyperspace vector $S_n(t)$ of the $2^M$ dimensional Hilbert space is generated by the bit string products of the reference noises corresponding to a number $n$, resulting in $2^M$ hyperspace vectors [12]. The number $n$, associated with the reference noises $\{G_1^n(t), G_2^n(t), \dots, G_M^n(t)\}$, is represented by the hyperspace vector:

$$S_n(t) = \prod_{i=1}^M G_i^n(t). \tag{4}$$

In squeezed INBL scheme, for each bit $i$, $G_i^n(t) = 1$ when the bit value is 0, and $G_i^n(t) = H_i(t)$ when the bit value is 1. Thus, the product strings need to only consider high bit values $H_i(t)$ which are RTWs, since the products of the low bit values $L_i(t)$ amount to multiplication with 1 [20]. This also means that the hyperspace vector for $n = 0$ is equal to constant 1. Following Eq. (4), hyperspace vectors $\{A(t), B(t), C(t) \dots\}$ can be constructed to represent an arbitrary set of numbers $\{A, B, C, \dots\}$, as seen below:

$$A(t) = G_1^A(t) \; G_2^A(t) \; G_3^A(t) \dots G_M^A(t) \tag{5a}$$
$$B(t) = G_1^B(t) \; G_2^B(t) \; G_3^B(t) \dots G_M^B(t) \tag{5b}$$
$$C(t) = G_1^C(t) \; G_2^C(t) \; G_3^C(t) \dots G_M^C(t) \tag{5c}$$
$$\vdots$$

Hyperspace vectors are themselves RTWs. For example, in the case of $M = 4$ noise-bits, the hyperspace vector for binary number $A = 1100$ is $A(t) = R_{11}(t)R_{21}(t).1.1 = R_{11}(t)R_{21}(t)$, see Figure 5. As demonstrated in this example, only the high noise bits contribute to the hyperspace composition.

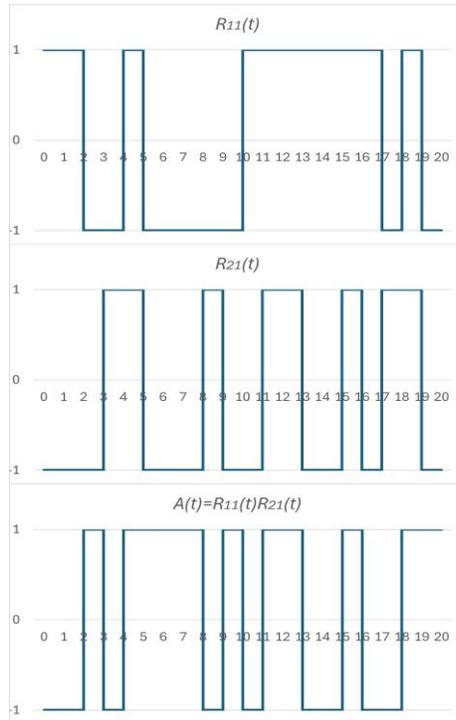

**Figure 5.** Illustration of the first 20 clock cycles for a given binary number $A = 1100$ represented by the hyperspace vector $A(t) = R_{11}(t)R_{21}(t)$.

### 2.3.   *Superpositions of Hyperspace Vectors and the Forming of the Universe*

Product strings such as hyperspace vectors can be superimposed. Sets of numbers in INBL are the superposition of their respective hyperspace vectors, enabling the transmission of large sets of numbers on a single wire without signal multiplexing, as well as parallel operations on these sets [14,16].

$$Y(t) = A(t) + B(t) + C(t) + \cdots \qquad (6)$$

For example, consider binary numbers as $A = 1100$, $B = 1010$ and $C = 1000$. Figure 6 demonstrates the superposition of the respective hyperspace vectors; that is, the sum $A(t) + B(t) + C(t)$.

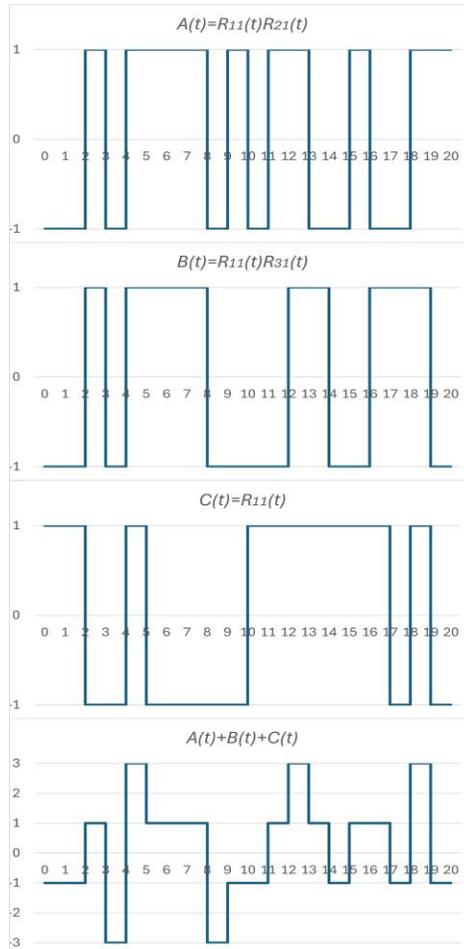

**Figure 6.** Illustration of the first 20 clock cycles for the case of $M = 4$ squeezed INBL scheme, given binary numbers $A = 1100$, $B = 1010$, and $C = 1000$, where the respective hyperspace vectors $A(t)$, $B(t)$, and $C(t)$ have been superimposed.

The superposition of all hyperspace vectors is called the universe, which can be generated with polynomial complexity through the Achilles' Heel algorithm modified for the squeezed INBL scheme [10,13], see Eq. (7). In this context, "Achilles Heel" refers to an explicit constructive procedure for building the complete superposition from the reference noises, rather than to a vulnerability.

$$U(t) = \sum_{k=1}^{2^M} S_k(t) = \prod_{i=1}^{M}[L_i(t) + H_i(t)] = \prod_{i=1}^{M}[1 + H_i(t)] \qquad (7)$$

This feature is similar to quantum computing where the Hadamard operation is used for this purpose. The analogy is at the level of state structure, to the action of tensor-product Hadamard gates in quantum computing, which map an initial computational basis state into an equal superposition over all $2^M$ basis states. In the squeezed INBL scheme, the Achilles Heel algorithm plays a similar role by generating a superposition ("universe") that contains all hyperspace vectors. However, the analogy is limited: INBL employs real-valued stochastic RTWs and classical addition of signals, so there is no underlying unitary evolution or complex phase structure as in genuine quantum Hadamard operations, and no nonclassical measurement statistics.

In the example of $M = 4$ noise-bits, the Achilles' Heel generated universe is shown on Eq. (8). Each factor in this equation is either 0 (for $R_{i1} = -1$) or 2 (for $R_{i1} = 1$). Thus, for the $M = 4$ universe, the amplitude can only be 0 or 16 with probability $P = (0.5)^4 = 0.0625$, as seen in Figure 7. For larger $M$ The amplitude of the Squeezed INBL universe tends towards zero. For this reason, the gate operations presented in this paper are limited to small superpositions only.

$$U(t) = [1 + R_{11}(t)][1 + R_{21}(t)][1 + R_{31}(t)][1 + R_{41}(t)] = 1 + \sum_{i=1}^{4} R_{i1} + \sum_{1 \le i < j \le 4} R_{i1}R_{j1} + \cdots \tag{8}$$

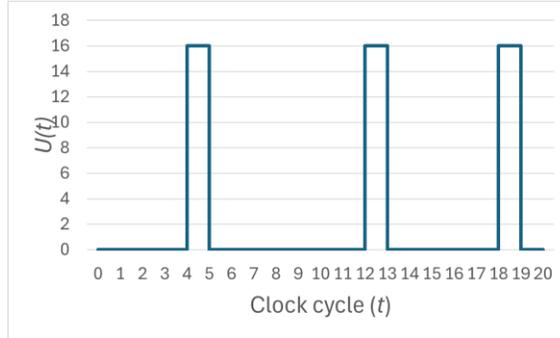

**Figure 7.** Illustration of the first 20 clock cycles for the $M = 4$ squeezed INBL universe $U(t)$.

### 2.4. *Review of the NOT Gate Implemented in Squeezed INBL*

Gate operations in INBL leverage the properties of the RTW reference noises described in the introduction: orthogonal Hadamard products and self-multiplication properties. Additionally, the distributive property of multiplication over addition enables any product operation applied to a superposition of hyperspace vectors to directly extend to each of the individual hyperspace vectors in the superposition. This enables inherently parallel computations to be executed simultaneously on all constituent hyperspace vectors within a superposition [5, 26, 27].

This section reviews the INBL NOT gate operation as a steppingstone, since the basic concepts for the NOT gate will lead naturally to the development of pairwise XOR and XNOR gate operations [21] for the squeezed INBL scheme. The NOT operation in [15] was developed for a reference system using symmetric RTWs, i.e., with amplitudes $\pm 1$. The NOT operation on the $x$-th noise-bit, illustrated in Figure 8, is defined as:

$$\text{NOT}_x(t) = R_{x0}(t)R_{x1}(t). \tag{9}$$

For squeezed INBL $R_{x0}(t) = 1$, so the NOT gate acting on the $x$-th noise-bit simplifies to:

$$\text{NOT}_x(t) = R_{x1}(t). \tag{10}$$

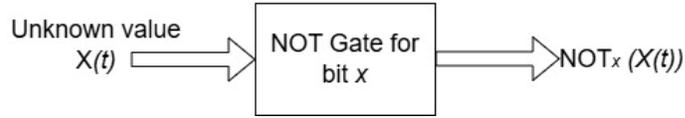

**Figure 8.** Implementation of the NOT gate operating on the $x$-th noise-bit.

For example, the RTW for $\text{NOT}_1(t) = R_{11}(t)$ operation targeting bit 1 within a product string shown in Figure 9, is as follows:

$$\text{NOT}_1(t)R_{10}(t) = [R_{10}(t)R_{11}(t)]R_{10}(t) = [(1)R_{11}(t)](1) = R_{11}(t) \tag{11}$$

or

$$\text{NOT}_1(t)R_{11}(t) = [R_{10}(t)R_{11}(t)]R_{11}(t) = [(1)R_{11}(t)]R_{11}(t) = 1. \tag{12}$$

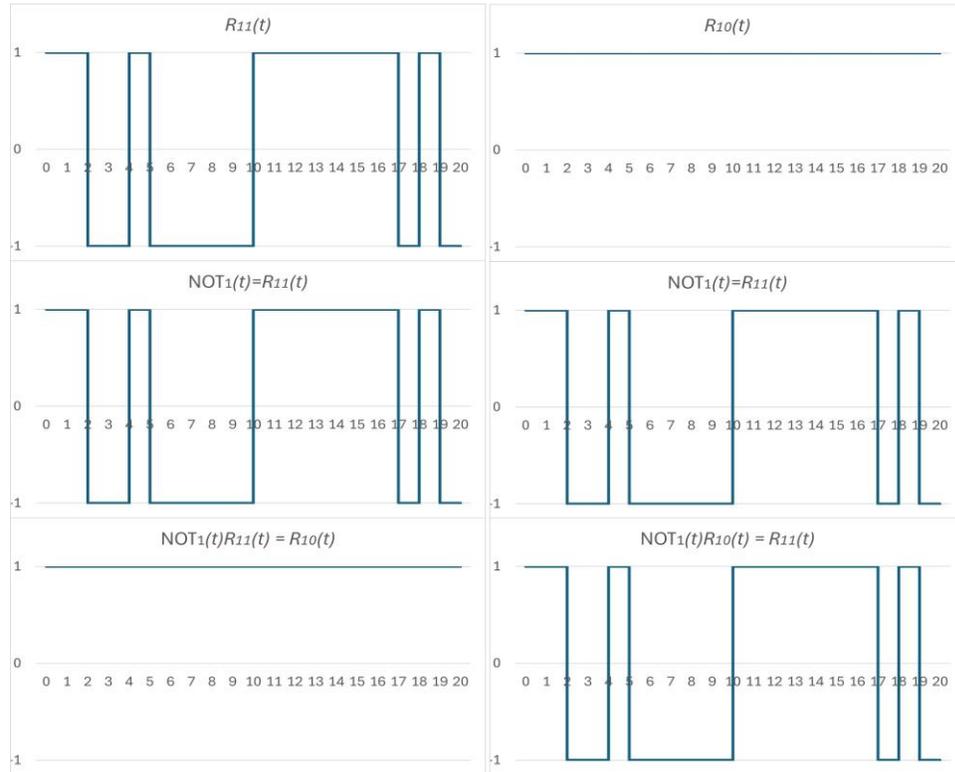

**Figure 9.** Illustration of the first 20 clock cycles for the $M = 4$ squeezed INBL $\text{NOT}_1(t)$ gate operation.

Due to the commutativity of multiplication, these relations work also on product strings, that is, on hyperspace vectors. Additionally, NOT operations in INBL can simultaneously target multiple bits. For example, bits 1 and 3 in bit string are inverted by $\text{NOT}_{13}$ operation:

$$\text{NOT}_{13}(t) = R_{10}(t)R_{11}(t)R_{30}(t)R_{31}(t) = R_{11}(t)R_{31}(t). \tag{13}$$

Figure 10 shows the $\text{NOT}_{13}$ operation on string $A = 1100$, represented by hyperspace vector $A(t)$, resulting in bit string $A_{13}^* = 0110$, represented by hyperspace vector $A_{13}^*(t)$:

$$A_{13}^*(t) = \text{NOT}_{13}(t)\text{A}(t) = R_{11}(t)R_{31}(t)R_{11}(t)R_{21}(t) = R_{21}(t)R_{31}(t). \tag{14}$$

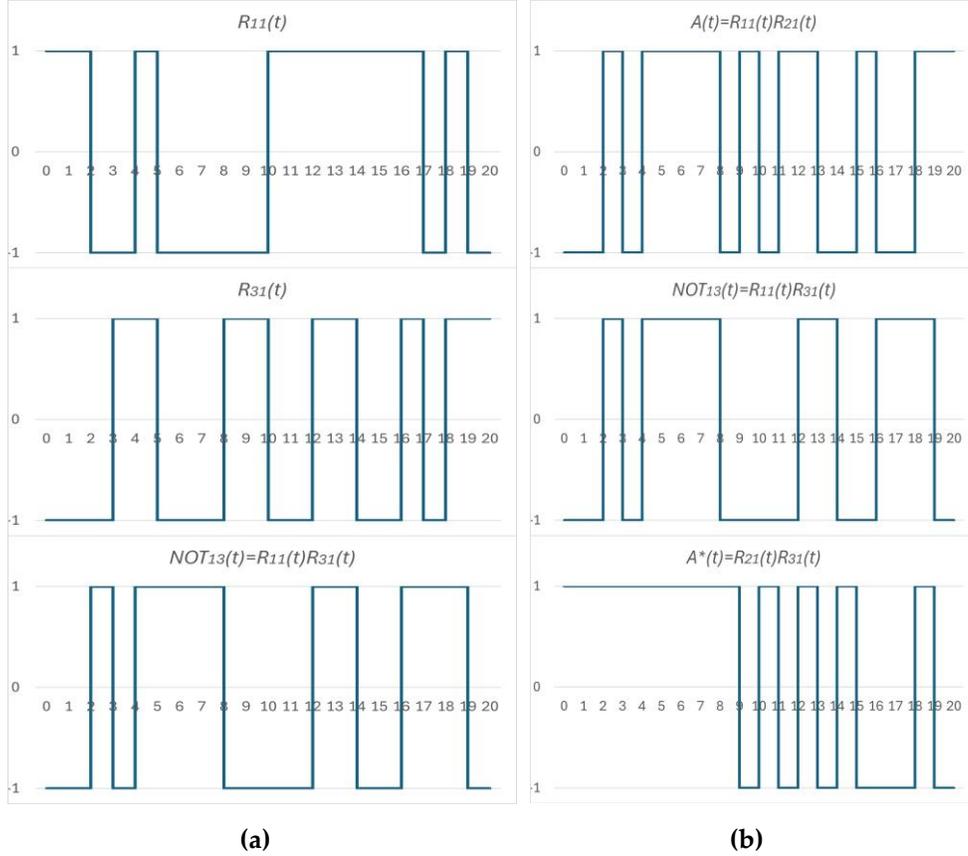

**(a)**                                       **(b)**

**Figure 10.** Example of NOT operation targeting multiple bits, showing the first 20 clock cycles for the $M = 4$ squeezed INBL: **(a)** $\text{NOT}_{13}(t) = R_{11}(t)R_{31}(t)$ operation will simultaneously invert bit 1 and bit 3 in a bit string (hyperspace vector) in a single operation; **(b)** For binary number $A = 1100$, bits 1 and 3 have been flipped simultaneously by $\text{NOT}_{13}$ operation acting on the respective hyperspace vector $A(t)$, resulting in $A_{13}^*(t)$ which corresponds to the binary number $A_{13}^* = 0110$.

The squeezed INBL NOT operation can be executed in parallel on sets of numbers represented by superpositions of hyperspace vectors. Consider the superposition of hyperspace vectors $A(t) = R_{11}(t)R_{21}(t)$, $B(t) = R_{11}(t)R_{31}(t)$ and $C(t) = R_{11}(t)$ shown in Figure 6. Implementing $\text{NOT}_{13} = R_{11}(t)R_{31}(t)$ on the superposition of hyperspace vectors representing set $\{1100, 1010, 1000\}$, results in another superposition of hyperspace vectors $A_{13}^*(t) = R_{21}(t)R_{31}(t)$, $B_{13}^* (t) = 1$ and $C_{13}^* = R_{31}(t)$

representing set $\{0110, 0000, 0010\}$; see Figure 11.

$$R_{11}(t)R_{31}(t)[R_{11}(t)R_{21}(t) + R_{11}(t)R_{31}(t) + R_{11}(t)] = R_{21}(t)R_{31}(t) + 1 + R_{31}(t)$$
(15)

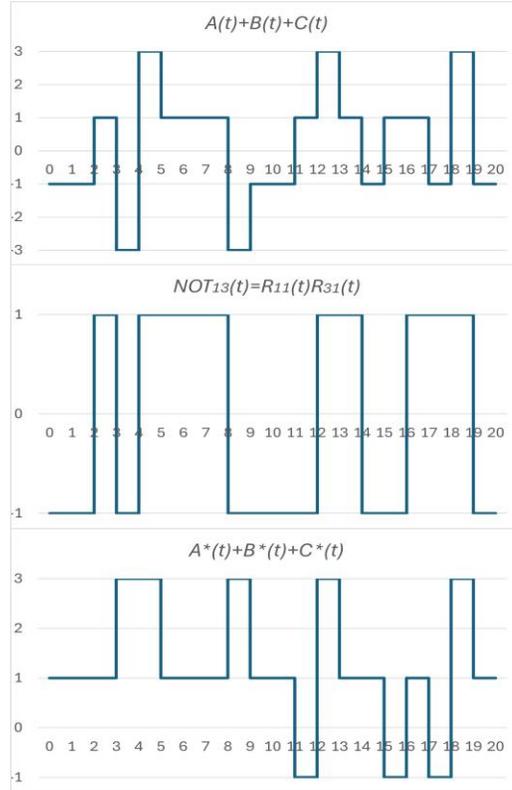

**Figure 11.** The first 20 clock cycles for the implementation of $\text{NOT}_{13}$ operation in $M = 4$ squeezed INBL. In this example, $\text{NOT}_{13}$ inverts bits $i = 1$ and $i = 3$ on a set of binary numbers represented by the superposition of hyperspace vectors $A(t)$, $B(t)$ and $C(t)$. The result is the superposition of hyperspace vectors $A^*_{13}$, $B^*_{13}$ and $C^*_{13}$. Note, the scale of amplitude axis of the superpositions is suppressed.

The NOT gate [15] described here highlights a powerful feature of INBL: simultaneous operations on bits can be carried out on superpositions at a constant time complexity regardless of how many numbers are included in the superposition. This is due to the distributive property of multiplication, where the NOT operation on hyperspace vectors can also be carried out on arbitrary superposition of hyperspace vectors. This review of the NOT gate implementation for the squeezed INBL scheme demonstrates the utilization of some of the properties that enable the pairwise XOR and XNOR gates that will be introduced in the next section.

## 3.   XOR and XNOR operations in Squeezed INBL

A previous work by Khreishah et al. [28] proposed bitwise XOR and XNOR operations implemented through manipulations of the reference wires, extending the concept originally introduced in [7,14]. However, altering the reference wires can disrupt the logic relationships implemented by other gates, which is a limitation for this approach [26]. In the Khreishah design, the XOR and XNOR gates take as inputs the bit values of two distinct noise-bits within a hyperspace vector, and the resulting output is obtained as the value of a third noise-bit within the same bit string.

A more recent work by Kenarangui et al. [21] introduced pairwise XOR and XNOR gate operations for the symmetric INBL scheme which does not require manipulation or "grounding" of the reference signals as in [7,13,14]. In this development, the inputs to these gate operations can be pairs of bits, hyperspace vectors, or superpositions. We build on this foundation by developing pairwise XOR and XNOR operations for the squeezed INBL scheme.

### 3.1. *Pairwise XOR and XNOR Operations on Noise-bits*

First consider XOR operation defined for a pair of two $i$-th noise-bits $G_i^j(t)$ and $G_i^k(t)$, where $j, k \in \{0, 1\}$. When $j \neq k$, the output of XOR operation must be logic high, and when $j = k$, the output of XOR operation must be logic low. Following from the properties of RTWs and the NOT gate operation introduced earlier (where $\text{NOT}_i(t) = G_i^1(t)$), the product of two noise-bits is as follows [21]:

$$G_i^j(t)G_i^k(t) = \begin{cases} G_i^0(t) = 1, & if \ j = k \\ \text{NOT}_i(t) = G_i^1(t), & if \ j \neq k \end{cases}.$$ (16)

The $\text{XOR}_i(t)$ gate operation on a pair of $i$-th noise-bits is accomplished through the product of RTWs with the low noise-bit $G_i^0(t) = 1$, where low valued noise-bits in squeezed INBL are equal to constant 1. See Figure 12 for demonstration of the XOR operation on the signals for the pair of $i = 1$ noise bits. We can see that for the XOR operation in squeezed INBL, multiplication by $G_i^0(t)$ has no impact.

$$\text{XOR}_i(t)[G_i^j(t)G_i^k(t)] = G_i^j(t)G_i^k(t)G_i^0(t) = \begin{cases} G_i^0(t) = 1, & if \ j = k \\ \text{NOT}_i(t)G_i^0(t) = G_i^1(t), & if \ j \neq k \end{cases}$$ (17)

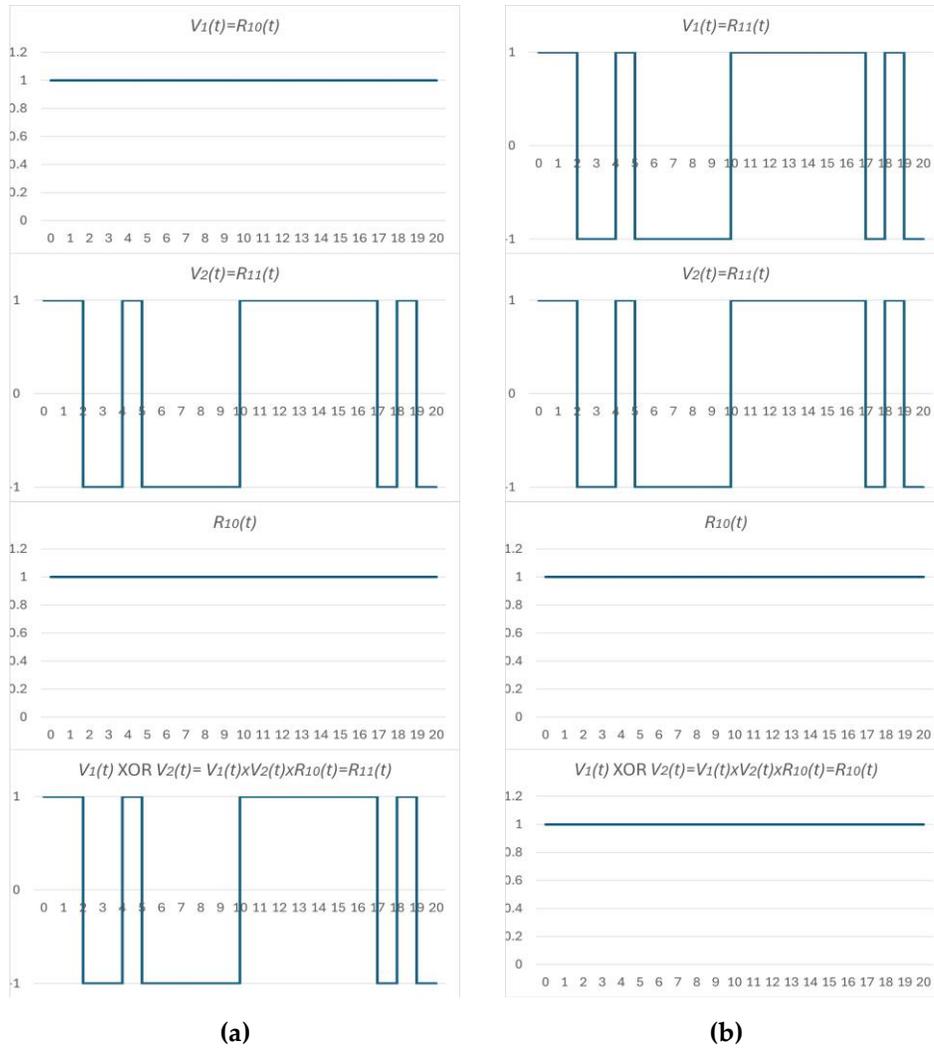

**(a)**                          **(b)**

**Figure 12.** Example of XOR operation on pairs of squeezed noise-bits, showing the first 20 clock cycles for the $M = 4$ squeezed INBL: **(a)** When noise-bit values j ≠ k, the output of XOR operation must be a high value $R_{11}(t)$ for noise bit 1; **(b)** when noise-bit values j = k, the output of XOR operation must be a low value $R_{10}(t)$ for noise bit 1.

The output of XNOR gate must be a low value, when $j \neq k$, and high value, when $j = k$. This operation is achieved in a similar way to the XOR gate, but instead through multiplication with the high noise-bit $G_i^1(t)$. See Figure 13 for demonstration of the XNOR operation on the signals for pair of 1-th noise bits.

$$\text{XNOR}_i(t)[G_i^j(t)G_i^k(t)] = G_i^j(t)G_i^k(t)G_i^1(t) =$$
$$\begin{cases} G_i^1(t), & if \ j = k \\ \text{NOT}_i(t)G_i^1(t) = \ G_i^0(t) = 1, & if \ j \neq k \end{cases} \qquad (18)$$

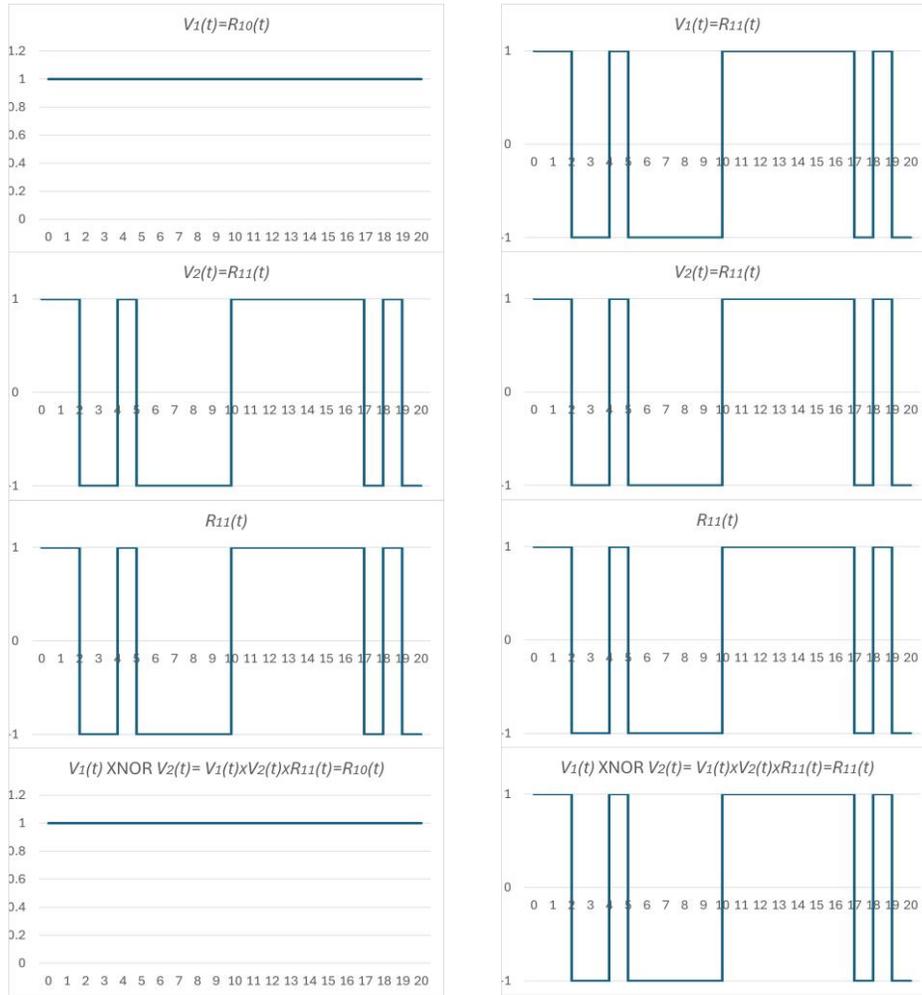

**(a)**          **(b)**

**Figure 13.** Example of XNOR operation on pairs of squeezed noise-bits, showing the first 20 clock cycles for the $M = 4$ squeezed INBL: **(a)** When noise-bit values j ≠ k, the output of XNOR operation must be a low value $R_{10}(t)$ for noise bit 1; **(b)** when noise-bit values j = k, the output of XNOR operation must be a high value $R_{11}(t)$ for noise bit 1.

It is demonstrated in this paper that XOR and XNOR gate operations can take a sum of noise-bits as one of the inputs, which would result in a sum of noise-bits as an output.

### 3.2. *Pairwise XOR and XNOR Operations on Bit Strings (Hyperspace Vectors)*

The pairwise XOR and XNOR operations, introduced for noise-bits in squeezed INBL, can be expanded for pairwise operations between bit strings or hyperspace vectors ($M$ bit long bit strings, where $M$ is the INBL bit resolution). Consider hyperspace vectors for two $M$ bit long bit strings $A$ and $B$:

$$A(t) \;=\; G_1^j(t)G_2^k(t)\;...\;G_M^l(t) \quad and \quad B(t) \;=\; G_1^m(t)G_2^n(t)\;...\;G_M^p(t). \tag{19}$$

Eq. (20) is the hyperspace vector for bit string of all low values in squeezed INBL, which is equal to constant 1. Using the property of operations implemented through multiplication, pairwise XOR operation between hyperspace vectors $A(t)$ and $B(t)$ is defined in Eq. (21).

$$Zeros(t) \;=\; G_1^0(t)G_2^0(t)\;...\;G_M^0(t) = 1 \times 1 \times ... = 1 \tag{20}$$

In the case of squeezed INBL, the multiplication by $Zeros(t) = 1$ has no impact on XOR gate operation. This simplifies the pairwise XOR operation into the Hadamard product between the hyperspace vectors $A(t)$ and $B(t)$, as seen in Eq. (22).

$$A(t)B(t)Zeros(t) = [G_1^j(t)G_1^m(t)G_1^0(t)]\,[G_2^k(t)G_2^n(t)G_2^0(t)]\,...\,[G_M^l(t)G_M^p(t)G_M^0(t)] \tag{21}$$

$$A(t)\mathrm{XOR}B(t) = A(t)B(t) = [G_1^j(t)G_1^m(t)]\,[G_2^k(t)G_2^n(t)]\,...\,[G_M^l(t)G_M^p(t)] \tag{22}$$

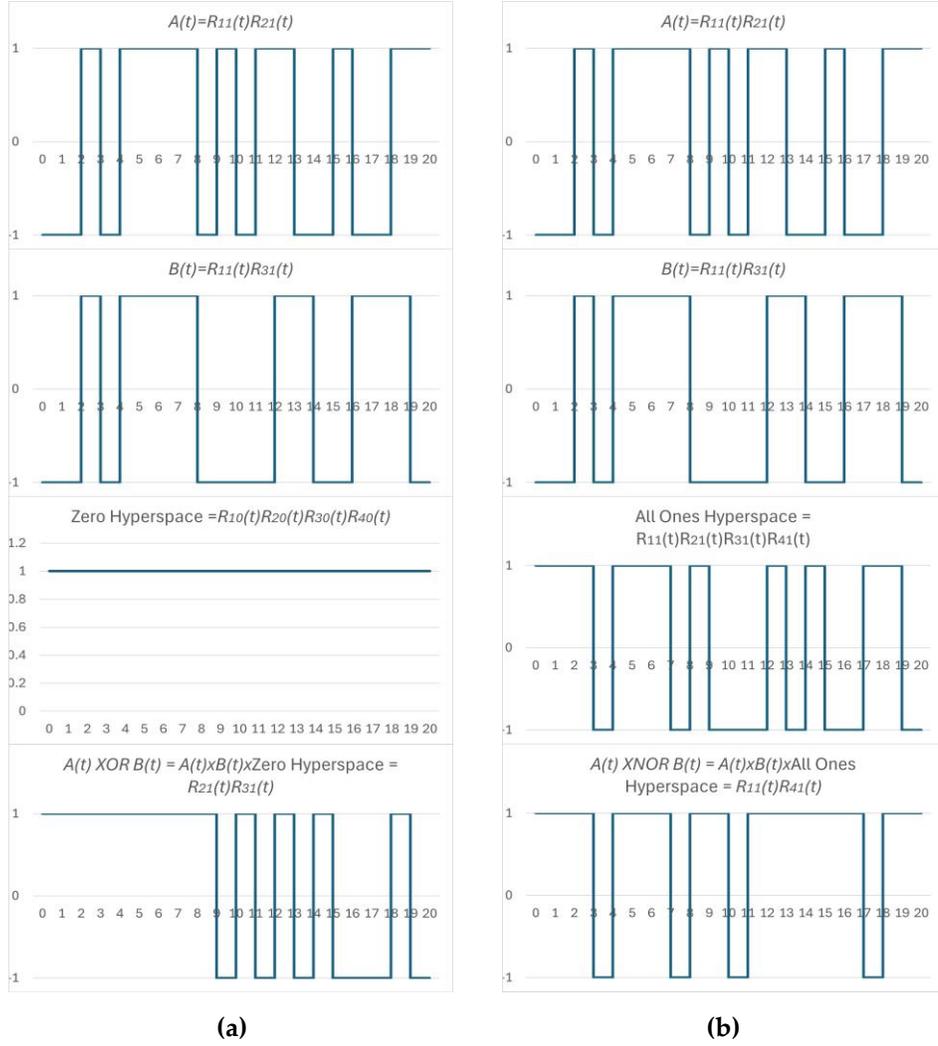

**(a)**                    **(b)**

**Figure 14.** Example of XOR and XNOR operations on a pair of hyperspace vectors $A(t)$ and $B(t)$, representing bit strings $A = 1100$ and $B = 1010$, showing the first 20 clock cycles for the $M = 4$ squeezed INBL: **(a)** Simultaneous pairwise XOR operation on hyperspace vectors, resulting in hyperspace vector representation of bit string $A$ XOR $B = 0110$ ; **(b)** simultaneous pairwise XNOR operation on hyperspace vectors, resulting in hyperspace vector representation of bit string $A$ XNOR $B = 1001$.

Eq. (23) is the hyperspace vector for bit string of all high values. The pairwise XNOR operation between hyperspace vectors $A(t)$ and $B(t)$ is defined in Eq. (24) [21].

$$Ones(t) = G_1^1(t)G_2^1(t) \ldots G_M^1(t) \tag{23}$$

$$A(t)B(t)Ones(t) = [G_1^j(t)G_1^n(t)G_1^1(t)][G_2^k(t)G_2^n(t)G_2^1(t)] \ldots [G_M^l(t)G_M^p(t)G_M^1(t)] \tag{24}$$

Consider the previous example for $M = 4$ bit squeezed INBL system, with bit strings $A = 0011$ and $B = 0101$, and their respective hyperspace vectors $A(t)$ and $B(t)$. The expected outputs for the pairwise XOR and XNOR gate operations are $A$ XOR $B = 0110$ and $A$ XNOR $B = 1001$, as demonstrated in Figure 14.

### 3.3. *Targeted XOR and XNOR Operations on Bit Strings (Hyperspace Vectors)*

Similar to the NOT operation in section 2, the XOR and XNOR operations described here can target specific bits in a bit string. For XOR, this is accomplished by setting the target noise-bits to the vector $G_{target}^p(t)$ which holds the value $p$, and setting the rest of the bits to their respective low noise-bit references $G_{i \neq target}^0(t)$. Consider for example hyperspace vector $A(t) = G_1^j(t) G_2^k(t) G_3^l(t) G_4^m(t)$ in $M = 4$ bit squeezed INBL system. We target $G_3^l(t)$, which is the third noise-bit with value $l$, by applying the XOR operation:

$$A(t) \text{ XOR}_3 \left[ G_1^0(t) G_2^0(t) G_3^p(t) G_4^0(t) \right] =$$
$$\left[ G_1^j(t) G_1^0(t) G_1^0(t) \right] \left[ G_2^k(t) G_2^0(t) G_2^0(t) \right] \left[ G_3^l(t) G_3^p(t) G_3^0(t) \right] \left[ G_4^m(t) G_4^0(t) G_4^0(t) \right]. \quad (25)$$

This is equivalent to simply pairwise XOR-ing between hyperspace vector $A(t)$ and only the third noise-bit $G_3^p(t)$. As shown previously, the XOR operation in squeezed INBL does not require multiplication by the term $G_3^0(t)$, whereas the symmetrical XOR operation does [21]. $A(t) \text{ XOR}_3 \ G_3^p(t)$ simplifies to $A(t) \ G_3^p(t)$, as follows:

$$A(t) \text{ XOR}_3 \ G_3^p(t) = \ G_1^j(t) \ G_2^k(t) [G_3^l(t) G_3^p(t)] G_4^m(t). \quad (26)$$

Similarly, for XNOR we set the target bits to the value $G_{target}^p(t)$, and set the rest of the bits to their respective high noise-bit $G_{i \neq target}^1(t)$. We target $G_3^l(t)$ by XNOR-ing it with only the third noise-bit $G_3^p(t)$ of value $p$. The $A(t) \text{ NXOR}_3 \ G_3^p(t)$ operation simplifies to Eq. (28):

$$A(t) \text{ XNOR}_3 \left[ G_1^0(t) G_2^0(t) G_3^p(t) G_4^0(t) \right] =$$
$$\left[ G_1^j(t) G_1^1(t) G_1^1(t) \right] \left[ G_2^k(t) G_2^1(t) G_2^1(t) \right] \left[ G_3^l(t) G_3^p(t) G_3^1(t) \right] \left[ G_4^m(t) G_4^1(t) G_4^1(t) \right]. \quad (27)$$

$$A(t) \text{ XNOR}_3 \ G_3^p(t) = \ G_1^j(t) \ G_2^k(t) [G_3^l(t) G_3^p(t) G_3^1(t)] G_4^m(t). \quad (28)$$

### 3.4. *Pairwise XOR and XNOR Operations on Superpositions*

Thus far it has been demonstrated that XOR and XNOR operations act pairwise between noise-bits, hyperspace vectors, or combinations of the two in order to target specific bits in a bit string, as seen in the previous subsection. These gate operations can be further expanded for superpositions. See Table 1 for summary of the INBL gate operations. While the inputs could be chosen to be any combination of noise-bits, hyperspace vectors, or superpositions, the pairwise XOR and XNOR operations on superpositions would have a more abstract application.

| NOT With a Hyperspace Vector | | NOT With a Superposition | |
|---|---|---|---|
| **Input** | **Output** | **Input** | **Output** |
| Hyperspace Vector | Hyperspace Vector | Superposition | Superposition |
| $A(t)$ | NOT(t) $A(t)$ | $A(t)$ | NOT(t) $A(t)$ |
| | | $B(t)$ | NOT(t) $B(t)$ |
| | | $C(t)$ | NOT(t) $C(t)$ |
| | | $\vdots$ | $\vdots$ |

| Pairwise XOR With a Hyperspace Vector | | Pairwise XOR With a Superposition | |
|---|---|---|---|
| **Inputs** | **Output** | **Inputs** | **Output** |
| Hyperspace vector | Hyperspace Vector | Hyperspace vector | Superposition |
| $X(t)$ | $X(t)$ XOR $A(t)$ | $X(t)$ | $X(t)$ XOR $A(t)$ |
| | | Superposition | $X(t)$ XOR $B(t)$ |
| Hyperspace vector | | $A(t)$ | $X(t)$ XOR $C(t)$ |
| | | $B(t)$ | $\vdots$ |
| $A(t)$ | | $C(t)$ | |
| | | $\vdots$ | |

| Pairwise XNOR With a Hyperspace Vector | | Pairwise XNOR With a Superposition | |
|---|---|---|---|
| **Inputs** | **Output** | **Inputs** | **Output** |
| Hyperspace vector | Hyperspace Vector | Hyperspace vector | Superposition |
| $X(t)$ | $X(t)$ XNOR $A(t)$ | $X(t)$ | $X(t)$ XNOR $A(t)$ |
| | | Superposition | $X(t)$ XNOR $B(t)$ |
| Hyperspace vector | | $A(t)$ | $X(t)$ XNOR $C(t)$ |
| | | $B(t)$ | $\vdots$ |
| $A(t)$ | | $C(t)$ | |
| | | $\vdots$ | |

**Table 1.** Examples of INBL gate operations, showing that the NOT operation is unary, while pairwise XOR and XNOR operations require two inputs [15,21]. The pairwise XOR and XNOR operations can take any combination of noise-bits, bit strings, and superpositions (all the different combinations not shown in figure).

The distributive property of multiplication guarantees that pairwise XOR and XNOR operations between bit strings of any size (including a single noise bit) can distribute over a sum of hyperspace vectors. We will demonstrate with a meaningful example, where one of the gate inputs is a hyperspace vector, while the other is a superposition. Consider the hyperspace vectors $A(t), B(t)$ and $C(t)$ discussed previously, which represent bit strings $A = 1100, B = 1010$, and $C = 1000$. In this example, XOR operation is implemented between hyperspace vector $A(t)$, and the superposition $B(t) + C(t)$:

$$A(t) \text{ XOR } [B(t) + C(t)] = A(t) \text{ XOR } B(t) + A(t) \text{ XOR } C(t) = R_{21}(t)R_{31}(t) + R_{21}(t). \quad (29)$$

Figure 15 (a) demonstrates the pairwise XOR operation between a hyperspace vector $A(t)$ and a superposition $B(t) + C(t)$. The output is confirmed to be the superposition of hyperspace vectors representing the set {0110, 0100}. Likewise, XNOR operation is implemented between hyperspace vector $A(t)$, and the superposition $B(t) + C(t)$:

$$A(t) \text{ XNOR } [B(t) + C(t)] = A(t) \text{ XNOR } B(t) + A(t) \text{ XNOR } C(t) = R_{11}R_{41} + R_{11}R_{31}R_{41}. \tag{30}$$

Figure 15 (b) demonstrates the pairwise XNOR $A(t)$ and a superposition $B(t) + C(t)$. The output is confirmed to be the superposition of hyperspace vectors representing the set {1001, 1011}.

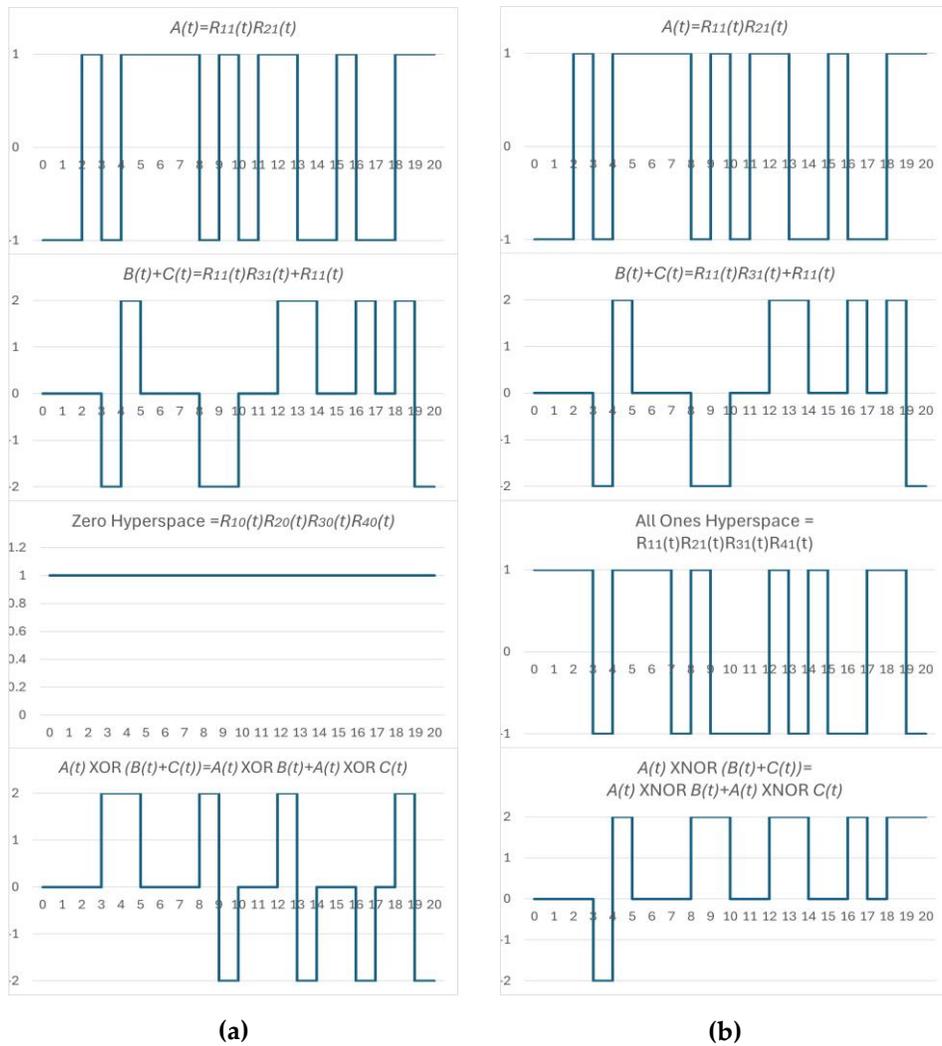

**(a)**                                        **(b)**

**Figure 15.** Example of XOR and XNOR operations between hyperspace vector $A(t)$ and superposition $B(t) +$ C(t), where bit strings $A = 1100$, $B = 1010$, and $C = 1000$, showing the first 20 clock cycles for the $M = 4$

squeezed INBL: **(a)** Pairwise XOR operation resulting in the output superposition representing the set{0110, 0100}; **(b)** Pairwise XNOR operation resulting in the output superposition representing the set {1001, 1011}.

## 4. Conclusions

RTWs in INBL can be realized as discrete-time stochastic waveforms generated by electronic circuits that approximate ideal random telegraph behavior, whether via PRNG-driven digital logic or via physical noise sources [22-24], subject to the orthogonality constraints discussed previously. The squeezed scheme reduces hardware overhead by mapping logic-low states to a constant waveform, so only the high-state RTWs require active generation and storage, which simplifies the reference noise system. The instantaneous nature of the gates reflects the fact that their implementation as algebraic products of RTWs does not require time averaging; instead, the relevant logic relations emerge at each clock cycle from the multiplicative properties and statistical independence of the reference noises.

The potential computational advantage of INBL over conventional Boolean logic arises from two features: first, hyperspace vectors and their superpositions allow exponentially many bit strings to be encoded in a single composite noise signal; second, gate operations implemented as multiplicative operations distribute over superpositions, enabling inherently parallel updates to entire sets of numbers in constant time. In this paper we verify that the squeezed XOR/XNOR gates inherit these structural properties: they operate instantaneously at the signal level and act consistently on noise-bits, hyperspace vectors, and superpositions. A full complexity comparison against standard Boolean implementations for specific tasks has been carried out previously for INBL in the asymmetric setting [5]. Reproducing such detailed algorithm-level comparisons for the squeezed gate set is left as future work.

It is useful to distinguish INBL from other stochastic or probabilistic computation paradigms, such as conventional stochastic computing and p-bit–based probabilistic computing [29,30]. In stochastic computing, real numbers are encoded as probabilities in bit streams and arithmetic is carried out via simple logic on these streams; correlations are generally a nuisance, and computations are typically serial or require careful decorrelation. In p-bit computing, probabilistic bits fluctuate between 0 and 1 with tunable bias and are coupled in networks to solve optimization or sampling problems. In contrast, INBL uses families of orthogonal RTWs as basis states to construct a high-dimensional hyperspace in which composite stochastic signals represent entire sets of bit strings, and logic gates act via multiplicative operations that distribute over superpositions. As a result, INBL's notion of "parallelism" and its use of orthogonality differ qualitatively from both stochastic computing and p-bit architectures. Beyond these structural distinctions, a key conceptual difference is that INBL is a deterministic computing framework implemented with stochastic carrier signals, rather than a stochastic computing paradigm in the sense of stochastic computing or p-bit networks. In many INBL algorithms, the only statistical aspect is that the correct output may not be available for a brief initial interval. However, (i) the probability that the output is not yet settled decays exponentially with waiting time, (ii) the eventual error probability is zero, and (iii) the system state itself makes it clear when the answer is not yet ready. This sharply contrasts with stochastic-computing and p-bit approaches, in which residual error and probabilistic outcomes are intrinsic features of the computation. A detailed quantitative comparison with these paradigms can be a topic of future research.

This work has developed pairwise XOR and XNOR gates for the squeezed INBL scheme. This is an extension of recent XOR and XNOR operations formulated for the symmetric reference system. The squeezed scheme maps the logic low states to a constant 1 while implementing RTWs only for the logic high states. This framework preserves the structure of the INBL system's implementation and operations while reducing the overhead resource requirements and simplifying the implementation of gate operations. Within this setting, the proposed pairwise XOR/XNOR gate definitions operate consistently at three levels: on individual noise-bits, on bit strings (including $M$-bit long hyperspace vectors), and on superpositions. In addition, it retains the instantaneous and distributive evaluation that enables parallel operations simultaneously on entire sets of numbers. The operations presented in this paper can be executed consecutively to accomplish even more complex data structure tasks. This can be a topic of future research.

These XOR and XNOR algorithms were originally inspired by the NOT operation previously developed for INBL, in which gate operation is implemented through a Hadamard product. They leverage the fact that when the reference noise-bits are equal $(j = k)$, their product forms a constant one, known as the "vacuum state" [20]. In contrast, when the noise-bits differ $(j \neq k)$, the product of the reference noises is the same as the NOT gate operation. One major implication for operations in the INBL system is when two noise-bit values are identical $(j = k)$, these algorithms cannot distinguish between the cases $(j = k = 0)$ and $(j = k = 1)$. This poses a challenge for the development of AND/OR operations for INBL using this simple Hadamard product-based execution. Development of AND/OR gate operations for INBL is a significant subject for future research, as it would help complete a universal gate set for INBL analogous to universal gate constructions in quantum computing.

Because the XOR and XNOR gate operations implemented in this paper avoid manipulation or "grounding" of reference wires, they circumvent a key drawback of earlier XOR/XNOR proposals whose wire-level modifications could interfere with other logic relations in the same reference system [26]. At the same time, the analysis of the squeezed "universe" shows that large superpositions tend to zero amplitudes. This confines the present gate operations to modest superposition sizes; addressing the zero-universe limitation is an important step for future research in INBL algorithm design.

More broadly, the availability of NOT, pairwise XOR, and pairwise XNOR in the squeezed INBL scheme strengthens the case for INBL as a flexible, classical platform capable of emulating features associated with quantum information processing, such as operations on superposed hyperspace vectors while relying only on classical stochastic signals.

## References

[1] S. Chen, "Quantum advantage showdowns have no clear winners," Wired (2022), https://www.wired.com/story/quantum-advantage-showdowns-have-no-clear-winners/ (accessed 15 January 2026).

[2] R. Brierley, "Not so fast," Nat. Phys. 17 (2021) 1073.

[3] E. Tang, "Quantum principal component analysis only achieves an exponential speedup because of its state preparation assumptions," Phys. Rev. Lett. 127 (2021) 060503.

[4] E. Tang, "Quantum-inspired classical algorithm for recommendation systems," in Proc. 51st Annu. ACM SIGACT Symp. Theory Comput. (STOC 2019), Phoenix, AZ, USA, 23–26 June 2019, pp. 217–228.

[5] E. Rieffel and W. Polak, Quantum Computing: A Gentle Introduction (MIT Press, Cambridge, MA, 2011).

[6] N. Kenarangui, W.C. Daugherity, A. Powalka and L.B. Kish, "Quantum supremacy challenged. Instantaneous noise-based logic with benchmark demonstrations," Fluct. Noise Lett. 24 (2025) 2540010.

[7] L.B. Kish and W.C. Daugherity, "Entanglement, and unsorted database search in noise-based logic," Appl. Sci. 9 (2019) 3029.

[8] L.B. Kish, "Noise-based logic: Binary, multi-valued, or fuzzy, with optional superposition of logic states," Phys. Lett. A 373 (2009) 911–918.

[9] S. Bezrukov and L.B. Kish, "Deterministic multivalued logic scheme for information processing and routing in the brain," Phys. Lett. A 373 (2009) 2338–2342.

[10] L.B. Kish, S. Khatri and S. Sethuraman, "Noise-based logic hyperspace with the superposition of 2^N states in a single wire," Phys. Lett. A 373 (2009) 1928–1934.

[11] L.B. Kish, S. Khatri and F. Peper, "Instantaneous noise-based logic," Fluct. Noise Lett. 9 (2010) 323–330.

[12] H. Wen and L.B. Kish, "Noise-based logic: Why noise? A comparative study of the necessity of randomness out of orthogonality," Fluct. Noise Lett. 11 (2012) 1250021.

[13] L.B. Kish, "Quantum supremacy revisited: Low-complexity, deterministic solutions of the original Deutsch–Jozsa problem in classical physical systems," R. Soc. Open Sci. 10 (2023) 221327.

[14] L.B. Kish and W.C. Daugherity, "Noise-based logic gates by operations on the reference system," Fluct. Noise Lett. 17 (2018) 1850033.

[15] F. Peper and L.B. Kish, "Instantaneous, non-squeezed, noise-based logic," Fluct. Noise Lett. 10 (2011) 231–237.

[16] H. Wen, L.B. Kish, A. Klappenecker and F. Peper, "New noise-based logic representations to avoid some problems with time complexity," Fluct. Noise Lett. 11 (2012) 1250003.

[17] M.A. Nielsen and I.L. Chuang, Quantum Computation and Quantum Information, 2nd ed. (Cambridge University Press, Cambridge, 2010) pp. 109–115.

[18] O. Cohen, "Classical teleportation of classical states," Fluct. Noise Lett. 6 (2006) C1–C8.

[19] O. Cohen, "Reply to 'On Cohen's classical teleportation of classical states'," Fluct. Noise Lett. 6 (2006) C31–C32.

[20] L.B. Kish, "Ternary noise-based logic," Fluct. Noise Lett. 23 (2024) 2450020.

[21] N. Kenarangui, A. Powalka and L.B. Kish, "New XOR and XNOR operations in instantaneous noise-based logic," Fluct. Noise Lett. (2025), https://arxiv.org/abs/2511.10679.

[22] F. Yu, L. Li, Q. Tang, S. Cai, Y. Song and Q. Xu, "A survey on true random number generators based on chaos," Discret. Dyn. Nat. Soc. 2019 (2019) 2545123.

[23] M. Park, J.V. Rodgers and D.P. Lathrop, "True random number generation using CMOS Boolean chaotic oscillators," Microelectron. J. 46 (2015) 1340–1347.

[24] J. Brown, J.F. Zhang, B. Zhou, M. Mehedi, P. Freitas and Z. Ji, "Random-telegraph-noise-enabled true random number generator for hardware security," Sci. Rep. 10 (2020) 17439.

[25] L.B. Kish, S. Khatri and T. Horvath, "Computation using noise-based logic: Efficient string verification over a slow communication channel," Eur. Phys. J. B 79 (2011) 85–90.

[26] W.C. Daugherity and L.B. Kish, "More on the reference-grounding-based search in noise-based logic," Fluct. Noise Lett. 21 (2022) 2250023.

[27] B. Zhang, L.B. Kish and C. Granqvist, "Drawing from hats by noise-based logic," Int. J. Parallel Emerg. Distrib. Syst. 32 (2017) 244–251.

[28] M.B. Khreishah, W.C. Daugherity and L.B. Kish, "XOR and XNOR gates in instantaneous noise-based logic," Fluct. Noise Lett. 22 (2023) 2350041.

[29] A. Alaghi and J.P. Hayes, "Survey of stochastic computing," ACM Trans. Embed. Comput. Syst. 12 (2013) 92.

[30] K.Y. Camsari, B.M. Sutton and S. Datta, "p-Bits for probabilistic spin logic," Phys. Rev. 9 (2019) 011045.